\documentclass[10pt,letterpaper,english,prb,twocolumn,superscriptaddress]{revtex4}
\usepackage{array}
\usepackage{amsmath}
\usepackage{graphicx}
\usepackage{amssymb}
\usepackage{latexsym}

\newcommand{\numSites}{\ensuremath{L}}
\newcommand{\numParticles}{\ensuremath{N}}
\newcommand{\numAuxiliary}{\ensuremath{m}}
\newcommand{\totalSpin}{\ensuremath{S}}
\newcommand{\numCorrelator}{\ensuremath{n}}
\renewcommand{\vec}[1]{\mathbf{#1}}

\usepackage{babel}

\begin{document}

\title{Approximating strongly correlated wave functions with correlator
product states}

\author{Hitesh J. Changlani}
\affiliation{Department of Chemistry and Chemical Biology, Cornell University,
Ithaca, New York 14853}
\affiliation{Department of Physics, Cornell University, Ithaca, New York 14853}

\author{Jesse M. Kinder}
\affiliation{Department of Chemistry and Chemical Biology, Cornell University,
Ithaca, New York 14853}

\author{C. J. Umrigar}
\affiliation{Department of Physics, Cornell University, Ithaca, New York 14853}

\author{Garnet Kin-Lic Chan}
\email{gc238@cornell.edu}
\affiliation{Department of Chemistry and Chemical Biology, Cornell University,
Ithaca, New York 14853}

\begin{abstract}
We describe correlator product states, a class of numerically efficient many-body
wave functions to describe strongly correlated wave functions
in any dimension.
Correlator product states introduce direct correlations between physical
degrees of freedom in a simple way, yet provide the flexibility to describe
a wide variety of systems. We show that many interesting wave functions can
be mapped exactly onto correlator product states, including Laughlin's quantum
Hall wave function, Kitaev's toric code states, and Huse and Elser's
frustrated spin states. We also outline the relationship between correlator
product states and other common families of variational wave functions
such as matrix product states, tensor product states, and resonating
valence bond states.
Variational calculations for the Heisenberg and
spinless Hubbard models demonstrate the promise  of correlator product
states for describing both two-dimensional and fermion correlations.
Even in one-dimensional systems, correlator product states
are competitive with matrix product states for a fixed number of variational parameters.
\end{abstract}

\date{\today}

\maketitle
How can one efficiently approximate an eigenstate of a strongly correlated
quantum system? In one-dimensional systems, the density matrix renormalization
group (DMRG) provides a powerful and systematic numerical approach.\cite{dmrg_white,dmrg_review}
However, the accuracy of the DMRG in two or more dimensions is limited
by the one-dimensional encoding of correlations in the matrix product
states (MPS) that form the variational basis of the DMRG.\cite{dmrg_review}
Generalizations of MPS to higher dimensions --- tensor network or
tensor product states (TPS)\cite{tps_nishino,tps_verstraete,tps_mera,tps_sandvik,tps_xiang,tps_wen}
--- have been introduced recently, but these engender considerable
computational complexity (which does \emph{not} arise with MPS). This
has made it difficult to practically extend the success and accuracy
of the DMRG to higher dimensions.

In this article we examine a different class of quantum states: \textit{correlator
product states} (CPS). Unlike MPS and TPS, which introduce auxiliary degrees of freedom
to generate correlations between physical degrees of freedom, CPS correlate the
physical degrees of freedom explicitly. The CPS form has been rediscovered many
times,\cite{nightingale,huse_elser,nishino_tvs} but the potential of
CPS as an alternative to MPS/TPS for systematically approximating
strongly correlated problems remains largely unexplored. Here we take
up this possibility. CPS share many of the local properties of MPS/TPS but
appear more suitable for practical calculations in more than one
dimension as well as for fermion systems. To establish the potential
of CPS, we analyze the relation between CPS and common families of
analytic and numerical trial wave functions. We then discuss the most
important properties of CPS: they permit efficient evaluation of observables
and efficient optimization. Finally, we present variational Monte
Carlo calculations for both spin and fermion systems. Our CPS results
compare favorably with calculations using other variational wave functions that contain
 a similar number of variational parameters.

\emph{Note:} As this manuscript was completed we were informed of
recent work by Isaev \textit{et al.~}on hierarchical mean-field theory
\cite{isaev} and by Mezzacapo \textit{et al.~}on entangled plaquette
states\cite{mezzacapo} as well as earlier work on string-bond states.\cite{stringbond}
All these studies consider wave functions similar to CPS and share
many of our own objectives. However, while our current efforts are
related, especially to Ref.~\onlinecite{mezzacapo}, we focus on
aspects of CPS not addressed in these other works, such as the relationship
with well-known analytical and numerical wave functions, and we consider
different physical problems, such as fermion simulations. Thus we
regard our work as complementary rather than overlapping.

\begin{figure}
\centerline{\includegraphics[width=\columnwidth]{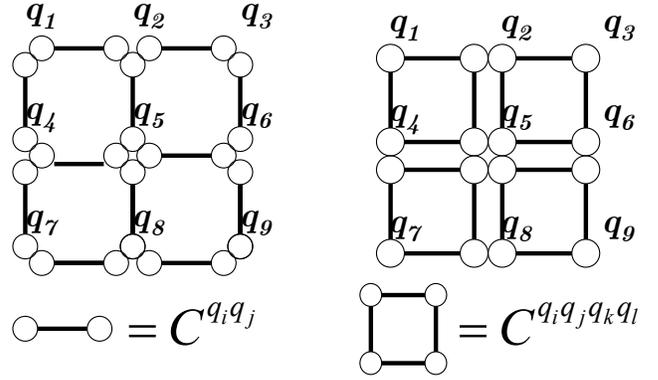}}

\caption{Nearest-neighbor 2-site and 2$\times$2 plaquette CPS on a 2D lattice.
The CPS weight for a given quantum configuration $|q_{1}q_{2}\ldots q_{\numSites}\rangle$
is obtained by multiplying correlator coefficients together as in
Eq.~(\ref{eq:nn_pair_cps}). \label{fig:cps_fig}}

\end{figure}

\section{Correlator Product States}

Consider a set of quantum degrees of freedom $Q\equiv\{q_{1},q_{2}\ldots q_{\numSites}\}$
on a lattice with \numSites{} sites in one or more dimensions. Each
$q_{i}$ might represent a spin $S=1/2$ degree of freedom, where $q \in \{\uparrow,\downarrow\}$,
or a fermion degree of freedom, in which case $q \in \{0,1\}$. An arbitrary
quantum state can be expanded over all configurations as
\begin{equation}
	|\Psi\rangle=\sum_{\{q\}}\Psi^{q_{1}q_{2}\ldots q_{\numSites}}|q_{1}q_{2}\ldots q_{\numSites}\rangle.\label{eq:full_ci}
\end{equation}
A general quantum wave function requires an exponential number of parameters ---
one for each configuration. One way to reduce the complexity of the problem is to
impose some structure on the coefficients $\Psi(Q)$.
Correlator product states (CPS) are one example of this approach.

CPS are obtained by associating variational
degrees of freedom directly with correlations between groups of sites.
For example, in the nearest-neighbor 2-site CPS, a correlator is associated
with each neighboring pair of sites:
\begin{equation}
	|\Psi\rangle =\sum_{\{q\}}\prod_{\langle ij\rangle}C^{q_{i}q_{j}}|q_{1}\ldots q_{\numSites}\rangle, \label{eq:nn_pair_cps}
\end{equation}
 where $\langle ij\rangle$ denotes nearest neighbors. The coefficients
in Eq.~(\ref{eq:full_ci}) are given by products of correlator coefficients.
For example, in a one-dimensional lattice, the amplitude of a configuration
is
\begin{equation}
	\Psi(Q)=C^{q_{1}q_{2}}C^{q_{2}q_{3}}C^{q_{3}q_{4}}\ldots C^{q_{\numSites-1}q_{\numSites}}.
\end{equation}
Eq.~(\ref{eq:nn_pair_cps}) can be extended to higher dimensions
simply by associating correlators with (overlapping) bonds on the
lattice (Fig.~\ref{fig:cps_fig}). The nearest-neighbor 2-site CPS
is an extremely simple CPS. Longer range correlations can be introduced
by removing the nearest neighbor restriction on pair correlations
or by including explicit correlations among more sites with correlators
such as $C^{q_{1}q_{2}q_{3}}$. It is clear that CPS provide a complete
basis: in the limit of $\numSites$-site correlators, the CPS amplitudes
are precisely the coefficients of Eq.~(\ref{eq:full_ci}).

When there is a global constraint on the total spin $\totalSpin{}$
or particle number $\numParticles{}$ we can use projected CPS wave
functions. For example, for fixed particle number, the $\numParticles$-projected
nearest-neighbor 2-site CPS is
\begin{equation}
	|\Psi\rangle=\sum_{\{q\}}\prod_{\langle ij\rangle}C^{q_{i}q_{j}}\hat{P}_{\numParticles}|q_{1}\ldots q_{\numSites}\rangle,
\end{equation}
where $\hat{P}_{\numParticles}$ ensures that $\sum_{i}q_{i}=N$. Such projections
do not introduce any complications in working with CPS and may be
included in both deterministic and stochastic calculations without difficulty.

It is sometimes useful to write the CPS in a different form. Each
correlator element $C^{q_{i}q_{j}}$ can be viewed as the matrix element
of a correlator operator $\hat{C}^{ij}$ that is diagonal in the quantum
basis $\{|q_{i}q_{j}\rangle\}$:
\begin{equation}
	\langle q_{i}q_{j}|\hat{C}^{ij}|q_{i}^{\prime}q_{j}^{\prime}\rangle=\delta_{q_{i}q_{i}\prime}\delta_{q_{j}q_{j}^{\prime}}C^{q_{i}q_{j}}.
\end{equation}
The CPS wave function is obtained by acting a string of commuting
correlator operators on a reference state $|\Phi\rangle$. For example,
a 2-site CPS may be written as
\begin{equation}
	|\Psi\rangle=\prod_{i>j}\hat{C}^{ij}|\Phi\rangle .\label{eq:cps2}
\end{equation}

When there are no constraints, the reference state is taken to be
an equally weighted sum over all quantum configurations:
$|\Phi\rangle=\sum_{\{q\}}|q_{1}q_{2}\ldots q_{\numSites}\rangle$;
otherwise, $|\Phi\rangle$ is projected to satisfy the constraint.
For example, if particle number is fixed, $|\Phi_{\numParticles}\rangle$ is an
equally weighted sum over all quantum configurations with particle
number $N$,
\begin{equation}
	|\Phi_{\numParticles}\rangle=\sum_{\{q\}}\hat{P}_{\numParticles}|q_{1}q_{2}\ldots q_{\numSites}\rangle.
\end{equation}
Note that both projectors and correlators are diagonal
operators in the Hilbert space and commute with one another: this means
that the projection can be applied directly to the reference state and this
simplifies  numerical algorithms using CPS. The operator
representation is also useful when considering extensions to the CPS
form such as alternative reference states.

\section{Connection to Other Wave Functions}

Many strongly correlated quantum states can be  represented exactly
as correlator product states. CPS also have much in common with several
classes of widely used variational wave functions: matrix product
states, tensor product states, and resonating valence bond states.
In this section, we discuss the connections between these wave functions.

\subsubsection*{Huse-Elser wave functions}

In their study of frustrated spin systems, Huse and Elser constructed
states in which the quantum amplitudes $\Psi(Q)$
correspond to classical Boltzmann weights $\exp(-\beta E[Q]/2)$ multiplied
by a complex phase.\cite{huse_elser} The weights are derived from
an effective classical Hamiltonian $\hat{H}^{cl}$. For example, in
the case of pairwise correlations, $\hat{H}^{cl}=\sum_{ij}\hat{h}_{ij}^{cl}$
with $\hat{h}_{ij}^{cl}=K_{ij}\hat{S}_{z}^{i}\hat{S}_{z}^{j}$. The
corresponding wave function can be represented as a 2-site CPS with
$\hat{C}^{ij}=\exp(-\beta\hat{h}_{ij}^{cl}/2+i\hat{\phi}_{ij})$ where
$\hat{\phi}_{ij}$ assigns a complex phase to the pair $ij$.

For the square and triangular Heisenberg lattices, Huse and Elser
demonstrated that a very compact variational ground-state could be
obtained with a semi-analytic three-parameter model for $\hat{H}^{cl}$
(containing up to 3-site interactions) and an analytically determined
phase. Although CPS can represent such highly constrained wave functions
for symmetric systems, it can also serve as the foundation of a more
general numerical method. By allowing correlators to vary freely and
by considering hierarchies of larger correlated clusters, we can hope
to construct rapidly converging approximations to \textit{arbitrary}
strongly correlated quantum states, as the DMRG does for one-dimensional
quantum problems.

\subsubsection*{Laughlin wave function}

In 1983, Laughlin proposed a variational wave function to
explain the fractional quantum Hall effect.\cite{laughlin} The Laughlin
wave function describes a strongly interacting system with topological
order. Like the Huse and Elser wave functions, Laughlin's wave function
can be associated with the Boltzmann weights of an  effective  classical Hamiltonian  and can be represented exactly
as a correlator product state.

The Laughlin quantum Hall state at filling fraction $1/m$ can be
written in first quantization as
\begin{equation}
	\Psi(\mathbf{r}_{1},\ldots\mathbf{r}_{\numParticles})=\prod_{\lambda<\mu}^{\numParticles}(z_{\lambda}-z_{\mu})^{m}e^{-\alpha\sum_{\kappa}|z_{\kappa}|^{2}}\label{eq:laughlin}
\end{equation}
where $z_{\lambda}$ is the (complex) coordinate of particle $\lambda$.
(A Greek subscript indicates the coordinate of a particular electron.
A Roman subscript indicates the coordinate of a lattice site.) Alternatively,
the system can be mapped onto a \textit{discrete} set of coordinates
$z_{1},\ldots,z_{\numSites}$ with an associated set of occupation numbers
$q_{1},\ldots,q_{\numSites}$. Then Eq.~(\ref{eq:laughlin}) can be exactly
expressed as a 2-site CPS in the occupation number representation
\begin{align}
	|\Psi\rangle & =\sum_{\{q\}}\prod_{i} C_{1}^{q_{i}}\prod_{i<j} C_{2}^{q_{i}q_{j}}\hat{P_{\numParticles}}|q_{1}\ldots q_{\numSites}\rangle\\
	\mathbf{C_{1}} & =\left(\begin{array}{c}
		1\\
		e^{-\alpha|z_{i}|^{2}}\end{array}\right)\\
	 \mathbf{C_{2}} & =\left(\begin{array}{cc}
		1 & 1\\
		1 & (z_{i}-z_{j})^{m}\end{array}\right)\label{eq:cps_laughlin}
\end{align}

The CPS wave function exactly reproduces the Laughlin wave function.
It is, in some ways, more general than Eq.~(\ref{eq:laughlin}).
The CPS form could be used to extend the Laughlin state beyond 2-site
correlators while maintaining antisymmetry of the state, or to find
a better variational energy in open or disordered systems.

\subsubsection*{Toric code}

Kitaev's toric code is another interesting quantum state with an exact
CPS representation. Kitaev proposed the toric code as a model for
topological quantum computing. The Hamiltonian is a sum of site and
plaquette projectors on a square lattice with spins placed on the
bonds. On a torus, the ground state of this Hamiltonian is 4-fold
degenerate with a gap to all other excitations.\cite{kitaev} It is
an example of a quantum system with topological order.

The ground state can be obtained from the zero-temperature Boltzmann
weights of a classical Hamiltonian
$\hat{H}_{\text{toric}}^{cl}=\sum_{\Box_{i}}\hat{h}_{\Box_{i}}$.
The sum is over all plaquettes $\Box_{i}$, and $\hat{h}_{\Box_{i}}$
is a product of $\hat{S}_{z}$ operators associated with the spins
on the edges of the plaquette.\cite{tps_verstraete} The amplitudes
of the toric code wave function can be generated by a CPS with plaquette
correlators:
\begin{equation}
	C_{\square}^{ijkl}=
	\begin{cases}
		1 & \mbox{if }S_{z}^{i}S_{z}^{j}S_{z}^{k}S_{z}^{l}>0,\\
		0 & \mbox{if }S_{z}^{i}S_{z}^{j}S_{z}^{k}S_{z}^{l}<0.
	\end{cases}
\end{equation}

The exact representation of the toric code and Laughlin's wave function
demonstrate the ability of CPS to describe systems with topological
order.

\subsubsection*{MPS and TPS}

Correlator product states are conceptually related to matrix
and tensor product states. All of these wave functions can easily
express entanglement between local degrees of freedom. Nonetheless,
CPS and MPS/TPS form different classes of quantum states and one is
not a proper subset of the other.

A matrix product state (MPS) is obtained by approximating the quantum
amplitudes $\Psi(Q)$ in Eq.~(\ref{eq:full_ci}) as a product of
matrices, one for each site on the lattice:

\begin{equation}
	\Psi(Q)=\sum_{\{i\}}{A}_{i_{1}i_{2}}^{q_{1}}{A}_{i_{2}i_{3}}^{q_{2}}\ldots A_{i_{\numSites}i_{1}}^{q_{\numSites}} \label{eq:mps}
\end{equation}
The {}``auxiliary'' indices $\{i\}$ are contracted in a one-dimensional
pattern --- a matrix product --- and this gives rise to the low
computational cost of working with MPS. However, the one-dimensional
structure prevents MPS from efficiently describing correlations in
higher dimensions. Tensor product states (TPS) extend MPS by approximating
the amplitudes $\Psi(Q)$ by more general tensor contractions.
Because of the more complicated contraction pattern, TPS can in principle
describe higher dimensional correlations.\cite{tps_verstraete,tps_sandvik,tps_xiang}
Unlike MPS, the TPS contraction \textit{cannot} be evaluated efficiently in general.
This leads to the high computational cost of working with TPS.

To demonstrate the relationship between CPS and MPS/TPS, we consider
a simple example of a nearest-neighbor 2-site CPS on a three-site
lattice with periodic boundary conditions. This CPS amplitudes are
\begin{equation}
	\Psi^{q_{1} q_{2} q_{3} } = C^{q_{1}q_{2}}C^{q_{2}q_{3}}C^{q_{3}q_{1}}
\end{equation}
Applying singular value decomposition to one of the correlators gives
\begin{equation}
	{C}^{qq^{\prime}}=\sum_{i}U_{i}^{q}\sigma_{i}V_{i}^{q^{\prime}}= \sum_{i} U_{i}^{q}W_{i}^{q^{\prime}},
\end{equation}
where we have absorbed the diagonal matrix $\sigma_{i}$ into $W_{i}^{q'}$.
With this decomposition, $|\Psi\rangle$ can be mapped to a MPS of
auxiliary dimension $2$:
\begin{equation}
	\Psi^{ q_{1} q_{2} q_{3} }=\sum_{\{i\}} U_{i_{1}}^{q_{1}} \left(W_{i_{1}}^{q_{2}}U_{i_{2}}^{q_{2}}\right)\left(W_{i_{2}}^{q_{3}}U_{i_{3}}^{q_{3}}\right) W_{i_{3}}^{q_{1}}.
\end{equation}
This is equivalent to Eq.~(\ref{eq:mps}) with $A_{ij}^{q}=W_{i}^{q}U_{j}^{q}$.
The matrices of the resulting MPS (of dimension $2$) have a restricted
form. More complicated CPS (e.g., with 3-site correlators) map to
MPS with larger auxiliary dimension and more flexible forms for the
matrices. (The dimension of the matrices grows exponentially with
the range or number of sites in the correlator.) An \textit{arbitrary}
MPS cannot be mapped onto a CPS with less than the complete basis
of $\numSites$-site correlators. Conversely, a one-dimensional CPS
with \textit{long-range} correlators (such as the general 2-site CPS
used in the Laughlin state) can only be represented by a MPS with
an auxiliary dimension that spans the full Hilbert space. These arguments
can be extended to higher dimensions and similar conclusions hold
for the mappings between CPS and TPS. For a given number of variational
degrees of freedom, only a subset of CPS can be exactly written as
MPS/TPS and vice versa.

While the correlators in the CPS have no auxiliary indices,
they could be augmented by additional auxiliary indices.
For example, string-bond states may be considered one-site correlators with
a pair of auxiliary indices.\cite{stringbond}
$\numCorrelator$-site correlators can be generalized in an analogous way.

The concept of an area law is sometimes used in the analysis of wave functions.
If the amount of entanglement between a system and its environment scales
with the area of the boundary between the two, the system is said to obey
an area law.
Arguments from quantum information theory suggest that wave functions that
satisfy an area law can accurately describe systems (in any dimension)
with a finite correlation length.\cite{area_law} (Some critical systems with
long-range correlations also satisfy an area law, but others may violate
the area law at zero temperature.)

MPS wave functions satisfy a one-dimensional area law
and have a finite correlation length.
(Long-range correlations can be reproduced
over a finite range, but they will eventually decay exponentially.)
TPS satisfy area laws in two or more dimensions.
CPS with local correlators like nearest neighbor pairs or plaquettes also
satisfy an area law, making them promising candidates for systems with a finite
correlation length. CPS with long-range correlators, such as those
used in the Laughlin wave function, are not constrained by an area law and can
describe even more entanglement between system and environment, obeying a
volume law instead.

\subsubsection*{RVB states}

Resonating valence bond (RVB) states are widely used in
strongly correlated quantum problems.\cite{rvb_anderson,rvb_sorella}
A fermion RVB state can be written as a product of a Jastrow factor
and a projected BCS wave function
\begin{equation}
	|\Psi_{\text{RVB}}\rangle=e^{\sum_{ij}J_{ij}\hat{n}_{i}\hat{n}_{j}}\, \hat{P}_{\numParticles}\, e^{\sum_{ij}\lambda_{ij}a_{i}^{\dag}a_{j}^{\dag}}|\text{vac}\rangle\label{eq:rvb}
\end{equation}
 where $J_{ij}$ and $\lambda_{ij}$ are commonly taken to be real.

There is a close relationship between CPS and RVB states. At half-filling,
the $N$-projected 2-site CPS can be expressed in the form of
Eq.~(\ref{eq:rvb}). Consider a dimer covering of the lattice. Let
$\lambda_{ij}=1$ for each pair $ij$ that is connected by a dimer
and $\lambda_{ij}=0$ otherwise. The corresponding projected BCS state
is the CPS reference $|\Phi_{\numParticles}\rangle$ defined earlier:
\begin{equation}
	\hat{P}_{\numParticles}\ e^{\sum_{i<j}a_{i}^{\dag}a_{j}^{\dag}}|\text{vac}\rangle=\hat{P}_{\numParticles}\sum_{\{q\}}|q_{1}q_{2}\ldots q_{\numSites}\rangle=|\Phi_{\numParticles}\rangle .
\end{equation}
 If the Jastrow factor is \textit{\emph{allowed to become }}\textit{complex},
then the 2-site correlator $\hat{C}^{ij}$ is fully parameterized as
\begin{equation}
	\hat{C}^{ij}=\exp(J_{0}+J_{1}^{i}\hat{n}_{i}+J_{1}^{j}\hat{n}_{j}+J_{2}^{ij}\hat{n}_{i}\hat{n}_{j}),
\end{equation}
and the CPS and RVB wave functions are identical.

Despite the existence of a mapping between the two wave functions,
the emphasis of the CPS parameterization is quite different from that
of commonly studied RVB states. For fermion RVB wave functions where
$J_{ij}$ is real, the Jastrow factor is positive and the nodes of
the fermion wave function are those of the projected BCS state. In
general, such a wave function cannot be exact. In contrast, the CPS
wave function can modify the nodes of the reference wave function
$|\Phi_{\numParticles}\rangle$ through the complex Jastrow factor. By using higher
order correlators, the CPS state can therefore become exact. While
the most flexible RVB/CPS form would combine a complex Jastrow factor
with an arbitrary projected BCS reference, there are computational
advantages to the simpler CPS reference, including the possibility
to efficiently evaluate observables without the use of a stochastic
algorithm.\cite{inprep}

\section{Computational Cost of CPS}

To be useful in practical calculations, a variational wave function
must allow \emph{efficient} evaluation of expectation values and optimization
of its parameters.

This combination of properties in matrix product states is responsible
for the success of the density matrix renormalization group. The expectation value
of typical observables can be evaluated exactly in a time which
is polynomial in the size of the system. Likewise, the amplitude of
a given configuration can also be evaluated exactly in polynomial
time. As shown in Eq.~(\ref{eq:mps}), the amplitude of a configuration
is the trace of the product of $\numSites$ independent $\numAuxiliary$$\times$$\numAuxiliary$
matrices, where $\numAuxiliary$ is the dimension of the auxiliary
indices $\{i\}$ and $\numSites$ is the number of lattice sites.
The cost for evaluating the amplitude is $\mathcal{O}(\numAuxiliary^{3}\numSites)$.

Tensor product states generalize the structure of MPS to higher dimensions,
but numerical efficiency is lost. In general, TPS amplitudes cannot
be evaluated exactly in polynomial time! Additional renormalization
procedures must be used while performing the contractions, which introduces
an error that depends on the system size. For fermions, such errors
can result in amplitudes or expectation values incompatible with a
fermion wave function as well as a variational energy below the fermion
ground state, a so-called $N$-representability problem. As a result,
only certain classes of TPS are capable of efficient polynomial simulation.

Like MPS, correlator product states allow efficient, exact evaluation
of wave function amplitudes and expectation values. For example, the
amplitudes of a pair CPS are
$\Psi(Q)=\prod_{i<j}C^{q_{i}q_{j}}.$
The amplitude is a \textit{simple product of numbers.} This is true
for any CPS, and thus the complexity is proportional only to the number
of correlators in the ansatz. This is manifestly polynomial in the
system size. In general, evaluation of the amplitude with \numCorrelator-site
correlators will require $\mathcal{O}(\numSites)$ multiplications
if the correlators act \emph{locally} --- e.g, nearest neighbors,
plaquettes, etc. --- and $\mathcal{O}(\numSites^{\numCorrelator})$
if there are no restrictions.

This property allows efficient Monte Carlo sampling
of expectation values. (Deterministic algorithms can also be used
but will be presented elsewhere.\cite{inprep}) Moreover, constraints
such as fixed particle number or total spin are easily handled within
the Monte Carlo algorithm by limiting the Metropolis walk to states
that satisfy these constraints. The expectation value of an operator
is given by $\langle A\rangle=\sum_{Q}P(Q) A(Q)$, where $P(Q)=|\Psi(Q)|^{2}$
and
\begin{equation}
	A(Q)=\sum_{Q'}\langle Q|\hat{A}|Q'\rangle \, \frac{\Psi(Q^{\prime})}{\Psi(Q)}.
\end{equation}
The sum over $Q'$ extends over only those $Q'$ for which
$\langle Q|\hat{A}|Q'\rangle \ne 0$.
As long as $\hat{A}$ is sparse in the chosen basis, its expectation value
can be evaluated efficiently.
If $|\Psi\rangle$ is local (e.g., nearest-neighbor pair CPS),
a further simplification occurs for operators such as
$a_{i}^{\dag} a_{j}$, $a_{i}^{\dag} a_{j}^{\dag} a_{k} a_{l}$, or $\vec{S}_i \cdot \vec{S}_j$.
For these operators,  most of the factors in $\Psi(Q)$ and $\Psi(Q')$
are identical and cancel from the ratio so that the
time required to evaluate the expecation value is independent of the system size
and depends only on the number of Monte Carlo samples.

As with MPS and TPS, we can take advantage of the product structure
of CPS when minimizing the variational energy and use an efficient
local optimization or ``sweep'' algorithm. The energy is minimized
with respect to one of the correlators while the others are held fixed,
then repeated for each of the correlators in turn until the energy
has converged. This algorithm is described in more detail in the next
section.

\section{\textit{\emph{Spin and Fermion Simulations}}}

We have implemented a pilot variational Monte Carlo code to optimize
general spin and fermion CPS wave functions. In Tables \ref{tab:heis}
and \ref{tab:hubbard} we present results for two models of interacting
spins and fermions: (i) the 2D square Heisenberg model defined by the Hamiltonian
\begin{equation}
	H=J\sum_{\langle ij\rangle}\mathbf{S}_{i}\cdot\mathbf{S}_{j},\label{eq:heisenberg}
\end{equation}
and (ii) a 1D \textit{\emph{spinless}} Hubbard model at half filling.
This model is defined by the Hamiltonian
\begin{equation}
	H=\sum_{\langle ij\rangle}-t(a_{i}^{\dag}a_{j}+a_{j}^{\dag}a_{i})+U n_{i} n_{j}.\label{eq:hubbard}
\end{equation}
Each site can only be occupied or unoccupied, and the energy $U$
is the cost of placing two fermions on neighboring sites.
We studied periodic and open boundary conditions for both the
Heisenberg and Hubbard models.

\begin{table*}

\caption{Variational Monte Carlo energies (in units of $J$) using CPS for the 2D $S=1/2$ Heisenberg
model, including percent errors ($\Delta E$). CPS{[}2{]} denotes nearest-neighbor
2-site correlators and CPS{[}$\numCorrelator$$\times$$\numCorrelator${]} denotes plaquette correlators.
The {}``exact'' $6$$\times$$6$ and $8$$\times$$8$ energies are obtained
from a stochastic series expansion MC calculation using \textsc{ALPS.}\cite{alps}
Unlike matrix product states, correlator product states maintain good
accuracy as the width is increased. \label{tab:heis}}

\begin{ruledtabular}
\begin{tabular}{clllllll}
Lattice & {CPS{[}2{]}} & $\Delta E$ & {CPS{[}$2$$\times$$2${]}} & $\Delta E$ & {CPS{[}$3$$\times$$3${]}} & $\Delta E$ & \multicolumn{1}{c}{Exact}\\
\hline
\multicolumn{8}{c}{\emph{Periodic Boundary Conditions}}\\
\hline
$4$$\times$$4$  & -11.057(1)  & 1.5\%  & -11.109(1)  & 1.1\%  & -11.2202(2) & 0.1\% & -11.2285\\
$6$$\times$$6$  & -23.816(3)  & 2.6\%  & -24.052(2)  & 1.6\%  & -24.313(2) & 0.5\% & -24.441(2) \\
$8$$\times$$8$  & -41.780(5)  & 3.1\%  & -42.338(4)  & 1.8\%  & -42.711(3) & 0.9\% & -43.105(3) \\
\hline
\multicolumn{8}{c}{\emph{Open Boundary Conditions}}\\
\hline
$4$$\times$$4$  & -8.8960(5)  & 3.2\%  & -9.0574(4)  & 1.4\%  & -9.1481(2) & 0.5\% & -9.1892\\
$6$$\times$$6$  & -20.811(1)  & 4.2\%  & -21.176(1)  & 2.5\%  & -21.510(1) & 1.0\% & -21.727(2) \\
$8$$\times$$8$  & -37.846(3)  & 4.5\%  & -38.511(2)  & 2.8\%  & -39.109(2) & 1.3\% & -39.616(2) \\
\end{tabular}
\end{ruledtabular}

\end{table*}

\begin{table*}

\caption{Variational Monte Carlo energies (in units of $t$) for the \emph{$L$}-site 1D spinless Hubbard
model with repulsion $U$ using periodic and open boundary conditions, including percent errors ($\Delta E$).
CPS{[}$n${]} denotes $n$-site correlators; DMRG{[}$m${]} denotes a DMRG calculation with
$m$ renormalised states. Since CPS and DMRG calculations are not
directly comparable in terms of complexity, the approximate number
of degrees of freedom per site (d.o.f.) is listed in the
bottom row. (The numbers are exact in the limit of an infinite
lattice.) Encouragingly, CPS are competitive with MPS for a comparable
number of variational parameters. {}Exact energies are from
$\numAuxiliary$=500 DMRG calculations. \label{tab:hubbard}}

\begin{ruledtabular}
\begin{tabular}{cc@{\hspace{12mm}}ll@{\hspace{12mm}}l@{\hspace{-15mm}}l@{\hspace{12mm}}ll@{\hspace{12mm}}l@{\hspace{-9mm}}l@{\hspace{12mm}}l}
\ $L$\  & \ $U$\ & \multicolumn{1}{c}{CPS{[}3{]}} & $\Delta E$ & \multicolumn{1}{c}{DMRG{[}3{]}} & $\Delta E$ & \multicolumn{1}{c}{CPS{[}4{]}} & $\Delta E$ & \multicolumn{1}{c}{DMRG{[}4{]}} & $\Delta E$ & \multicolumn{1}{c}{Exact}\\
\hline
\multicolumn{11}{c}{\emph{Periodic Boundary Conditions}}\\
\hline
12 & 0 & -7.052(1)	& 5.5\%	& -7.165	& 4.0\%	& -7.213(1)		& 3.4\%	& -7.313	& 2.0\%	& -7.464	\\
12 & 4 & -2.692(2)	& 4.1\%	& -2.577	& 8.2\%	& -2.756(1)		& 1.8\%	& -2.725	& 2.9\%	& -2.807	\\
12 & 8 & -1.461(1)	& 1.1\%	& -1.430	& 3.1\%	& -1.474(1)		& 0.2\%	& -1.462	& 1.0\%	& -1.477	\\
\hline
24 & 0 & -14.432(2)	& 5.0\%	& -14.608	& 3.8\%	& -14.714(2)	& 3.2\%	& -14.832	& 2.4\%	& -15.192	\\
24 & 4 & -5.34(1)	& 5.1\%	& -5.340	& 5.1\%	& -5.482(1)		& 2.6\%	& -5.403	& 4.0\%	& -5.626	\\
24 & 8 & -2.929(2)  & 0.8\%	& -2.860	& 3.2\%	& -2.931(1)		& 0.7\%	& -2.900	& 1.8\%	& -2.953	\\
\hline
36 & 0 & -21.82(1)	& 4.6\%	& -22.035	& 3.6\%	& -22.21(1)		& 2.8\%	& -22.421	& 1.9\%	& -22.860	\\
36 & 4 & -7.93(3)	& 6.0\%	& -8.127	& 3.7\%	& -8.17(1)		& 3.2\%	& -8.173	& 3.2\%	& -8.440	\\
36 & 8 & -4.390(2)	& 0.9\%	& -4.302	& 2.9\%	& -4.400(1)		& 0.7\%	& -4.355	& 1.7\%	& -4.430	\\
\hline
\multicolumn{11}{c}{\emph{Open Boundary Conditions}}\\
\hline
12 & 0 & -7.204(1)	& 1.3\%	& -7.185	& 1.5\%	& -7.274(1)		& 0.3\%	& -7.265	& 0.4\%	& -7.296	\\
12 & 4 & -3.748(1)	& 4.0\%	& -3.787	& 3.0\%	& -3.887(1)		& 0.5\%	& -3.894	& 0.3\%	& -3.905	\\
12 & 8 & -2.847(2)	& 4.6\%	& -2.920	& 2.2\%	& -2.971(1)		& 0.4\%	& -2.981	& 0.1\%	& -2.984	\\
\hline
24 & 0 & -14.593(1)	& 2.2\%	& -14.609	& 2.1\%	& -14.767(1)	& 1.1\%	& -14.838	& 0.6\%	& -14.926	\\
24 & 4 & -6.32(1)	& 7.8\%	& -6.543	& 4.5\%	& -6.687(1)		& 2.4\%	& -6.803	& 0.7\%	& -6.851	\\
24 & 8 & -4.287(2)	& 6.6\%	& -4.414	& 3.8\%	& -4.498(2)		& 2.0\%	& -4.576	& 0.3\%	& -4.590	\\
\hline
36 & 0 & -21.978(2)	& 2.6\%	& -22.035	& 2.3\%	& -22.260(2)	& 1.3\%	& -22.421	& 0.6\%	& -22.562	\\
36 & 4 & -8.83(3)	& 9.1\%	& -9.323	& 4.0\%	& -9.36(1)		& 3.6\%	& -9.625	& 0.9\%	& -9.713	\\
36 & 8 & -5.660(2)	& 7.3\%	& -5.873	& 3.8\%	& -5.934(3)		& 2.8\%	& -6.078	& 0.4\%	& -6.104	\\
\hline
\multicolumn{2}{c}{d.o.f} & \multicolumn{2}{l}{\quad 8} &
\multicolumn{2}{l}{\quad 18} & \multicolumn{2}{l}{\quad 16} &
\multicolumn{2}{l}{\quad 32} & \\
\end{tabular}
\end{ruledtabular}

\end{table*}

\subsubsection*{Optimization method}

We optimize the correlators by minimizing the variational energy
with a sweep algorithm.
At each step of each sweep, a target correlator is updated
while the other correlators are fixed. Because
the wave function $|\Psi\rangle$ is linear in the target correlator
coefficients, the derivatives of $|\Psi\rangle$ with respect
to these coefficients define a vector space for the optimization.
For instance, if the target correlator has elements $C^{\mu}$,
then the vector space is generated by the basis $|\tilde{\Psi}_{\mu}\rangle$
where
\begin{equation}
	|\tilde{\Psi}_{\mu}\rangle=\dfrac{\partial |\Psi\rangle}{\partial C^{\mu}}.
\end{equation}
Any vector in this space defines a CPS wave function: $\vec{x}$ corresponds to
the wave function $|\Psi(\vec{x})\rangle = /sum_{\mu} x^\mu |\tilde{\Psi}_\mu \rangle$.

It is convenient to work in a slightly different basis in which one vector
$\vec{x}_0$ corresponds to the current value of the target correlator
and the other vectors $\vec{x}_i$ are orthogonal to $\vec{x}_0$
(but not necessarily to each other).
The updated target correlator will be a linear combination of the
$\vec{x}_\alpha$ where $\alpha \in \{0,i\}$.

We construct the Hamiltonian $\mathcal{H}_{\alpha\beta}$ and the
overlap matrix $\mathcal{S}_{\alpha\beta}$ in this space:
\begin{align}
	\mathcal{H}_{\alpha\beta}= & \langle
\Psi(\vec{x}_{\alpha})|\hat{H}|\Psi(\vec{x}_{\beta})\rangle\\
	\mathcal{S}_{\alpha\beta}= & \langle
\Psi(\vec{x}_{\alpha})|\Psi(\vec{x}_{\beta})\rangle,
\end{align}
where $\hat{H}$ is the model Hamiltonian of Eq.~(\ref{eq:heisenberg})
or (\ref{eq:hubbard}). We then solve the generalized eigenvalue
problem
\begin{equation}
	\mathcal{H}\cdot\vec{C}=\lambda\mathcal{S}\cdot\vec{C},
\end{equation}
where $\vec{C}$ is a linear combination of the $\vec{x}_\alpha$.
The eigenvector with the lowest eigenvalue defines the optimal target correlator coefficients
$\tilde{C}^{\mu}$ that give the lowest energy when all other
correlators are fixed. We sweep over all of the correlators one at
a time until the energy stops decreasing.

This defines a general sweep algorithm for optimizing CPS. However to converge
the sweeps when  the Hamiltonian and overlap matrix are constructed via Monte Carlo sampling
it is very important to minimize the stochastic error.
Nightingale and Melik-Alaverdian,\cite{nightingale_opt} and Toulouse and Umrigar,\cite{umrigar_toulouse}
defined efficient estimators for variational Monte Carlo optimization,
and we have used these to construct $\mathcal{H}$ and $\mathcal{S}$.
For numerical stability, it is important to monitor the change in the variational parameters
and reject extremely large changes during a single iteration.\cite{umrigar_toulouse}
For CPS, this can be achieved by adding a dynamically adjusted diagonal shift to $\mathcal{H}$ that
penalizes large changes away from $C^\mu$: $\delta\mathcal{H}_{00} = 0$,
$\delta\mathcal{H}_{ii} > 0$.
Using this sweep algorithm, we find that the variational energy of the CPS converges
(within statistical error) in less than 5 sweeps.

To obtain the numbers in Tables \ref{tab:heis} and \ref{tab:hubbard},
we ran the linear optimization routine for each system through 3 or 4 sweeps,
after which the energy stopped decreasing and instead fluctuated within
a small range of values.
We chose one wavefunction (set of correlators) from the final sweep and
calculated the energy and variance reported in the tables
using a larger number of Monte Carlo samples
than we used during the optimization procedure.

\subsubsection*{Results}

Table~\ref{tab:heis} shows the optimized energies obtained for
the 2D square Heisenberg model. This model tests the ability
of CPS to describe two-dimensional correlations. When open boundary
conditions are used, the system is not translation invariant and
requires the kind of general parameterization of the CPS emphasized
here rather than the more restricted forms used by Huse and Elser.\cite{huse_elser}

The nearest-neighbor 2-site CPS (CPS{[}2{]}) has only four variational
parameters per site and gives errors in the range of 3--5\% for
open boundary conditions and 1--3\% for periodic boundary conditions.
The error is rapidly reduced by increasing the correlator size.
For example, for the 8$\times$8 periodic model,
going from pair to $2$$\times$$2$ to $3$$\times$$ 3$ plaquettes, the error
goes from 3.1\% to 1.8\% to 0.9\%. The rapid convergence
of the error with the correlator size is consistent with the
results of Mezzacapo \emph{et al.}~for hardcore boson systems with periodic
boundary conditions.\cite{mezzacapo}

As discussed earlier, CPS with local correlators like those used in
Table~\ref{tab:heis} satisfy an area law,
which allows them to accurately simulate systems with a finite correlation length.
However, the 2D Heisenberg model is gapless with long-range correlations,
so we expect the error to increase as the size of the lattice increases.
Nonetheless, the energetic error of the CPS wave function with a fixed
correlator size grows quite slowly as the lattice size is increased.
This is not true of MPS, in which the number of variational
parameters per site required to achieve a given accuracy grows rapidly
with the width of a 2D system.

We performed a series of DMRG calculations for the Heisenberg model on
the lattices in Table~\ref{tab:heis} with a range of values of $\numAuxiliary$
using \textsc{ALPS.}\cite{alps}
The variational objects in the DMRG are $\numAuxiliary $$\times$$ \numAuxiliary$ matrices.
For periodic boundary conditions, $\numAuxiliary \approx 35$, 250, and 750 are required
for 1 percent accuracy on the $4$$\times$$4$, $6$$\times$$6$, and $8$$\times$$8$ lattices
respectively. The latter calculation, which utilizes
about 1.1 million variational parameters per site
(neglecting symmetry and conservation of quantum numbers),
is to be contrasted with the much more compact
description using the CPS with $3$$\times$$3$ correlators, which corresponds to
just 512 parameters per site.

The spinless 1D Hubbard model with periodic boundary conditions has
nontrivial fermion correlations and cannot be mapped onto a local
spin model. Consequently, this model tests the ability of the CPS
to capture fermion correlations. In Table~\ref{tab:hubbard} we compare
3-site and 4-site nearest-neighbor CPS energies (CPS{[}3{]} and CPS{[}4{]})
with DMRG calculations for $\numAuxiliary=3$ and $\numAuxiliary=4$
renormalised states.
DMRG calculations were carried out using \textsc{ALPS.}\cite{alps}
For open boundary conditions, the error in the CPS energy is smallest in the noninteracting
system and largest for an intermediate interaction strength ($U=4$).
For periodic boundary conditions, the CPS{[}4{]} errors range from less than
1\% for the $U$=8 case to approximately 3\% for the free fermion
system --- a difficult limit for a locally entangled state.
The DMRG energies follow the same trends.

To make a meaningful comparison with the DMRG results, we also show
the approximate number of variational degrees of freedom per site
in each ansatz. A DMRG{[}$\numAuxiliary${]} wave function has
$\mathcal{O}(2\numAuxiliary^{2}\numSites)$
degrees of freedom (2 $\numAuxiliary$$\times$$\numAuxiliary$ matrices
at each site) whereas the CPS{[}$\numCorrelator${]} wave function
has $\mathcal{O}(2^{\numCorrelator}\numSites)$ degrees of freedom
(an $\numCorrelator$-site correlator at each site).
As a result, the CPS{[}4{]} wave function has a similar complexity
to the DMRG{[}3{]} state.
Depending on the boundary conditions and the length of the lattice, the
exact number of degrees of freedom may be less than this estimate.
For instance, when $\numSites=12$ for an open chain, the DMRG{[}3{]}
wave function has about 14.7 parameters per site and the CPS{[}4{]}
wave function has 12.
Comparing the CPS and DMRG calculations with similar numbers of variational
parameters, we see that the CPS energies are indeed very competitive,
especially for periodic boundary conditions, where a CPS includes
direct correlations between the ends of the chain.

Minimizing the CPS energy is a nonlinear optimization problem and the sweep
algorithm may not converge to the global minimum of the variational energy.
We have repeated the optimization
for different initial wave functions to avoid local minima.
The DMRG algorithm can also converge to a local minimum for $\numAuxiliary = 3$ or 4.
We repeated each of these DMRG calculation 100 times with the same
input and reported the lowest energy obtained in Table~\ref{tab:hubbard}.
Although convergence to local minima is possible in both CPS and DMRG calculations,
we believe the results reported in Tables~\ref{tab:heis} and \ref{tab:hubbard}
indicate the competitive accuracy of CPS as a general variational method.

\section{Conclusion}

In this paper, we evaluated correlator product states as a route
to describing strongly correlated wave functions in any dimension.
Our preliminary numerical studies indicate that CPS can capture both nontrivial
fermion correlations and two-dimensional correlations. Together with
the analysis showing the connections between CPS and many
interesting quantum states, this supports the intriguing possibility
that CPS are sufficiently flexible to systematically approximate general
strongly correlated spin and fermion problems in two or more dimensions.

Nonetheless, many questions remain to be answered. For example, how well do CPS
reproduce correlation functions? While properties are harder to obtain
accurately than energies in variational calculations, our view is
that so long as successive CPS[\numCorrelator] calculations form a sufficiently
rapidly convergent approximation to the quantum state of interest,
then accurate approximations to correlation functions can be constructed,
as in the case of DMRG calculations. Detailed investigations of such
questions and the analysis of more complex systems like the full Hubbard
model or the \emph{t-J} model will require more sophisticated numerical
treatments and alternative numerical techniques such as deterministic
evaluation methods. We are currently exploring these areas.

We thank C.L.~Henley for bringing the work of Huse and Elser to our
attention, T.~Nishino for pointing out the long history of CPS, and
S.R.~White for helpful discussions.

This work was supported by the National Science Foundation through CHE-0645380,
the DOE-CMSN program through DE-FG02-07ER46365,
the David and Lucile Packard Foundation,
the Alfred P. Sloan Foundation,
and the Camille and Henry Dreyfus Foundation.

\bibliography{cps_prb}
\end{document}